\documentclass[twocolumn]{aastex62}

\usepackage{placeins}
\usepackage{natbib}
\usepackage{amsmath}
\usepackage{color}
\usepackage{graphicx}
\usepackage{enumerate}

\newcommand\msun{M_\odot}

\shorttitle{Diverse Protoplanetary Disk Morphology} 
\shortauthors{Bae et al.}

\begin{document}

\title{DIVERSE PROTOPLANETARY DISK MORPHOLOGY PRODUCED BY A JUPITER-MASS PLANET}

\author[0000-0001-7258-770X]{Jaehan Bae}
\affil{Department of Terrestrial Magnetism, Carnegie Institution for Science, 5241 Broad Branch Road, NW, Washington, DC 20015, USA}

\author[0000-0001-8764-1780]{Paola Pinilla}
\affil{Department of Astronomy/Steward Observatory, The University of Arizona, 933 North Cherry Avenue, Tucson, AZ 85721, USA}

\author[0000-0002-1899-8783]{Tilman Birnstiel}
\affil{University Observatory, Faculty of Physics, Ludwig-Maximilians-Universit\"{a}t M\"{u}nchen, Scheinerstr. 1, D-81679 Munich, Germany}

\correspondingauthor{Jaehan Bae}
\email{jbae@carnegiescience.edu}

\begin{abstract}

Combining hydrodynamic planet-disk interaction simulations with dust evolution models, we show that protoplanetary disks having a giant planet can reveal diverse morphology in (sub-)millimeter continuum, including a full disk without significant radial structure, a transition disk with an inner cavity, a disk with a single gap and a central continuum peak, and a disk with multiple rings and gaps.
Such a diversity originates from (1) the level of viscous transport in the disk which determines the number of gaps a planet can open; (2) the size and spatial distributions of grains determined by the coagulation, fragmentation, and radial drift, which in turn affects the emmisivity of the disk at (sub-)millimeter wavelengths; and (3) the angular resolution used to observe the disk. 
In particular, our results show that disks having the same underlying gas distribution can have very different grain size/spatial distributions and thus appearance in continuum, depending on the interplay among coagulation, fragmentation, and radial drift.
This suggests that proper treatments for the grain growth have to be included in models of protoplanetary disks concerning continuum properties and that complementary molecular line observations are highly desired in addition to continuum observations to reveal the true nature of disks.
The fact that a single planet can produce diverse disk morphology emphasizes the need to search for more direct, localized signatures of planets in order to confirm (or dispute) the planetary origin of observed ringed substructures.
\end{abstract}


\section{Introduction}

\begin{figure*}
\centering
  \includegraphics[width=0.95\textwidth]{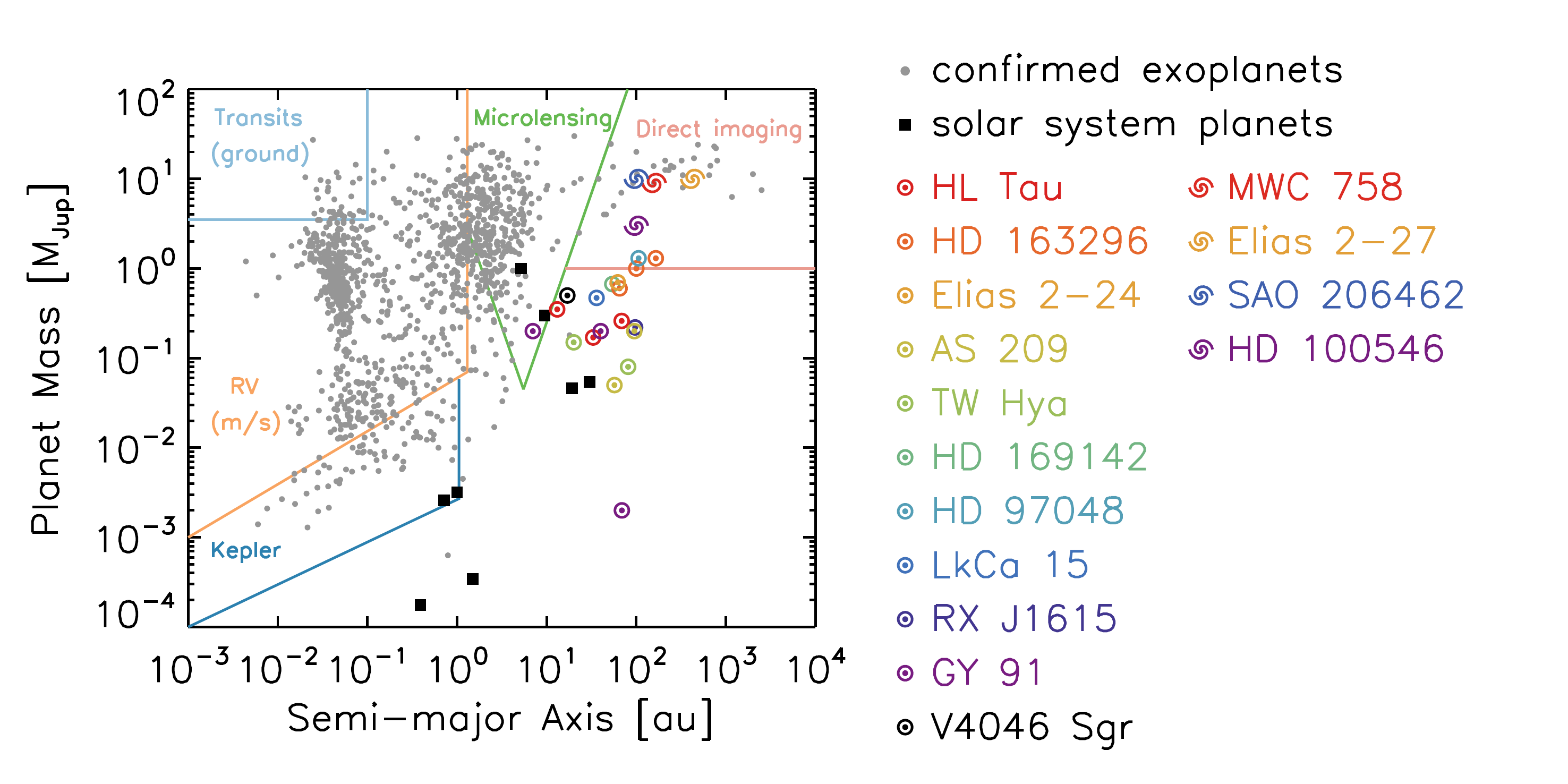}
\caption{The semi-major axis and mass of hypothesized planets assumed to reproduce the observed gaps in protoplanetary disks ($\odot$ symbols), compiled from literature: HL~Tau \citep{jin16}, HD~163296 \citep{teague18}, Elias~2-24 \citep{dipierro18}, AS~209 \citep{fedele18}, TW~Hya, HD~169142, HD~97048, LkCa~15, RX~J1615 \citep{dongfung17}, GY~91 \citep{sheehan18}, and V4046~Sgr (Ru\'iz-Rodr\'iguez et al. in prep.). The predictions are made using either planet-disk interaction simulations (HL~Tau, HD~163296, Elias~2-24, AS~209, V4046~Sgr) or empirical relations between planet's mass and gap width (TW~Hya, HD~169142, HD~97048, LkCa~15, RX~J1615, GY~91). For the estimations collected from \citet{dongfung17}, we adopted the masses obtained with a disk viscosity of $\alpha=10^{-3}$. Using a factor of 10 larger/smaller disk viscosity would result in about a factor of 3 larger/smaller planet masses \citep{dongfung17}. Also shown with spiral symbols are the semi-major axis and mass of hypothesized planets needed to reproduce observed spiral arms, based on planet-disk interaction simulations: MWC~758 \citep{dong15}, Elias~2-27 \citep{meru17}, SAO~206462 \citep{bae16c}, and HD~100546 \citep{follette17}. The gray circles present confirmed exoplanets as of 2018 May (\url{https://exoplanetarchive.ipac.caltech.edu/}). The black squares present the eight solar system planets. The over-plotted lines with color show illustrative estimates of the regions for which various exoplanet detection techniques have discovered exoplanets, similar to the shaded regions in Figure 6 of \citet{gaudi12}.}
\label{fig:substructure}
\end{figure*}

Followed by the revolutionary discovery of sets of rings and gaps in the millimeter continuum emission of the HL Tau disk \citep{alma15}, concentric rings and gaps have been imaged in more than a dozen of protoplanetary disks by now thanks to the Atacama Large Millimetre Array and optical/infrared telescopes equipped with adaptive optics.
Such ringed substructures are found in disks around stars with a broad range of masses from sub-solar masses (e.g., TW~Hya; \citealt{andrews16,tsukagoshi16}) to $\sim 2$ solar masses (e.g., HD~163296; \citealt{isella16}), but also with various ages spanning from less than 1 million years old (e.g., GY~91; \citealt{sheehan18}) to nearly 10 million years old (e.g., TW~Hya; \citealt{andrews16,tsukagoshi16}).
Furthermore, ringed substructures have been observed using different techniques: (sub-)millimeter/centimeter continuum \citep[e.g.,][]{alma15}, molecular line emission \citep{teague18}, and optical/infrared scattered light \citep[e.g.,][]{avenhaus18}.
These observations thus seem to suggest that ringed substructures are pervasive on scales of $0''.1$ ($\simeq 10 - 20$~au in linear scale) in protoplanetary disks \citep{zhang16,avenhaus18}.

The origin of observed rings and gaps is unfortunately still far from clear.
The interaction between planet and disk \citep{lin80} is certainly an intriguing possibility but other processes including various types of fluid instabilities \citep{takahashi14,flock17,dullemond18}, dust property changes across condensation fronts \citep{zhang15,okuzumi16}, and radial variation of magnetic activities \citep{johansen09} can also produce similar ringed substructures, and we do not have a conclusive way yet to differentiate these mechanisms based on observed features.

The situation became more complicated as it is shown that one planet can open multiple gaps (\citealt{bae17}; see also \citealt{dong17}). 
Planets excite multiple spiral arms \citep{baezhu18} and each spiral arm can open a gap at the radial location it shocks the disk gas \citep{bae17}.
In such a case, even with arbitrarily powerful observing facilities we would not detect a planet in the planet-induced gaps other than the primary one, paradoxically.

Assuming that an observed gap is created by a planet orbiting within them, one can estimate the planet's mass using hydrodynamic simulations or empirical relations between planet mass and gap depth/width \citep[e.g.,][]{kanagawa15}.
Figure \ref{fig:substructure} presents such attempts complied from literature, in which we plot the mass and semi-major axis of hypothesized planets required to reproduce the observed gaps.
The planet masses obtained by both approaches (i.e., simulations, empirical relations) depend on the physical properties of the underlying disk, including the disk aspect ratio and the level of viscous transport, but it is interesting to note that the estimated masses are broadly consistent for the 11 disks presented in the figure (18 gaps in total).
The estimated masses range from about a few percent of a Jupiter-mass to one Jupiter-mass, coincident with the masses of solar system's gas/ice giants.
If (and only if) confirmed, these gap-opening protoplanets will provide us critical insights into studies of planet formation.
Also, depending on their future migration and accretion we may be witnessing planets that will eventually be mini-Neptune-mass planets or hot/warm Jupiters, for which we now have a decent number of confirmed population in mature planetary systems.
As an aside, it is also worth mentioning that the planet masses assumed to reproduce observed gaps are approximately an order of magnitude smaller than the ones needed to reproduce observed spiral arms, presumably because the spiral arms driven by (sub-)Jovian-mass planets are too tightly wound and/or do not produce sufficient perturbations \citep{dongfung17b,baezhu18b}.

In this paper, as a step forward to better understand the origin of observed ringed substructures, we examine the morphology of protoplanetary disks in millimeter continuum produced by a Jupiter-mass planet.
Since we are concerned with millimeter continuum, we consider grain evolution -- both spatial and size -- in response to the gas structure that a Jupiter-mass planet creates.
As we will show, depending on the physical properties of the disk and the angular resolution used for observations, a single giant planet can produce a diverse disk morphology in millimeter continuum emission: (1) a full disk without significant radial structure; (2) a transition disk with an inner cavity; (3) a disk with a single gap with a central continuum peak; and (4) a disk with multiple rings and gaps.
From the perspective of differentiating possible gap-opening mechanisms, the diverse morphology a planet can produce emphasizes the need to search for more direct evidence of planets to confirm (or dispute) the planetary origin of observed ringed substructures.

This paper is organized as follows.
We introduce our hydrodynamic and dust evolutionary models in Section \ref{sec:methods}.
We present simulation results in Section \ref{sec:results} and  synthesized images of disks' continuum emission in Section \ref{sec:morphology}.
We summarize and present an outlook for future studies in Section \ref{sec:summary}.

\section{Methods and Background}
\label{sec:methods}

\subsection{Gas Evolution}

We run two-dimensional hydrodynamic calculations using FARGO3D \citep{benitez16,masset00} to simulate the gas evolution.
The initial disk assumes power-law gas surface density and temperature distributions following $\Sigma_{\rm gas} (r) = \Sigma_{{\rm gas},p} (r/ r_p )^{-1}$ and $T(r) = T_p (r/ r_p )^{-1/2}$, where $\Sigma_{{\rm gas},p}$ and $T_p$ are the gas surface density and temperature at the location of the planet $r=r_p$.
The simulation domain extends from 2 to 200~au in radius and from 0 to $2\pi$ in azimuth.
We adopt 2048 logarithmically spaced grid cells in the radial direction and 2792 uniformly spaced grid cells in the azimuthal direction.
We include a 1 $M_{\rm Jup}$ planet orbiting around a 1 $\msun$ star at $r_p=20$~au.
We assume a total disk mass of $0.02~\msun$ and a disk aspect ratio of 0.1 at $r=r_p$, with which $\Sigma_{{\rm gas},p} = 7.1~{\rm g~cm^{-2}}$ and $T_p = 127~{\rm K}$, respectively. 
We add a viscosity $\alpha=5\times10^{-5}$ or $5 \times 10^{-4}$ to the gas (see Section \ref{sec:models}), where $\alpha$ denotes a Shakura-Sunyaev viscosity parameter \citep{shakura73}.

\subsection{Grain Evolution}

For the spatial and size evolution of grains, we adopt the model introduced in \citet{birnstiel10} which we briefly summarize here.
In this model grains evolve in response to the underlying gas structure, considering growth via coagulation, fragmentation due to turbulence, and radial drift.
In practice, the grain evolution is simulated by solving the continuity equation for each radial grid cell and for each grain size bin.
We adopt the azimuthally averaged gas surface density distribution from hydrodynamic simulations obtained after 1000 planetary orbital times as the initial condition, by which time the disk has reached a quasi-steady state.
The initial grain surface density is constructed assuming a uniform gas-to-dust ratio of 100 throughout the disk, and we add the entire solid mass ($0.0002~\msun$) in $1~\mu$m grains.
We use 180 logarithmically spaced bins for grain sizes, ranging from $1~\mu$m to 2~m.
The grain evolution model does not include the back reaction of grains on to the gas.

\begin{figure*}[t]
\centering
  \includegraphics[width=0.95\textwidth]{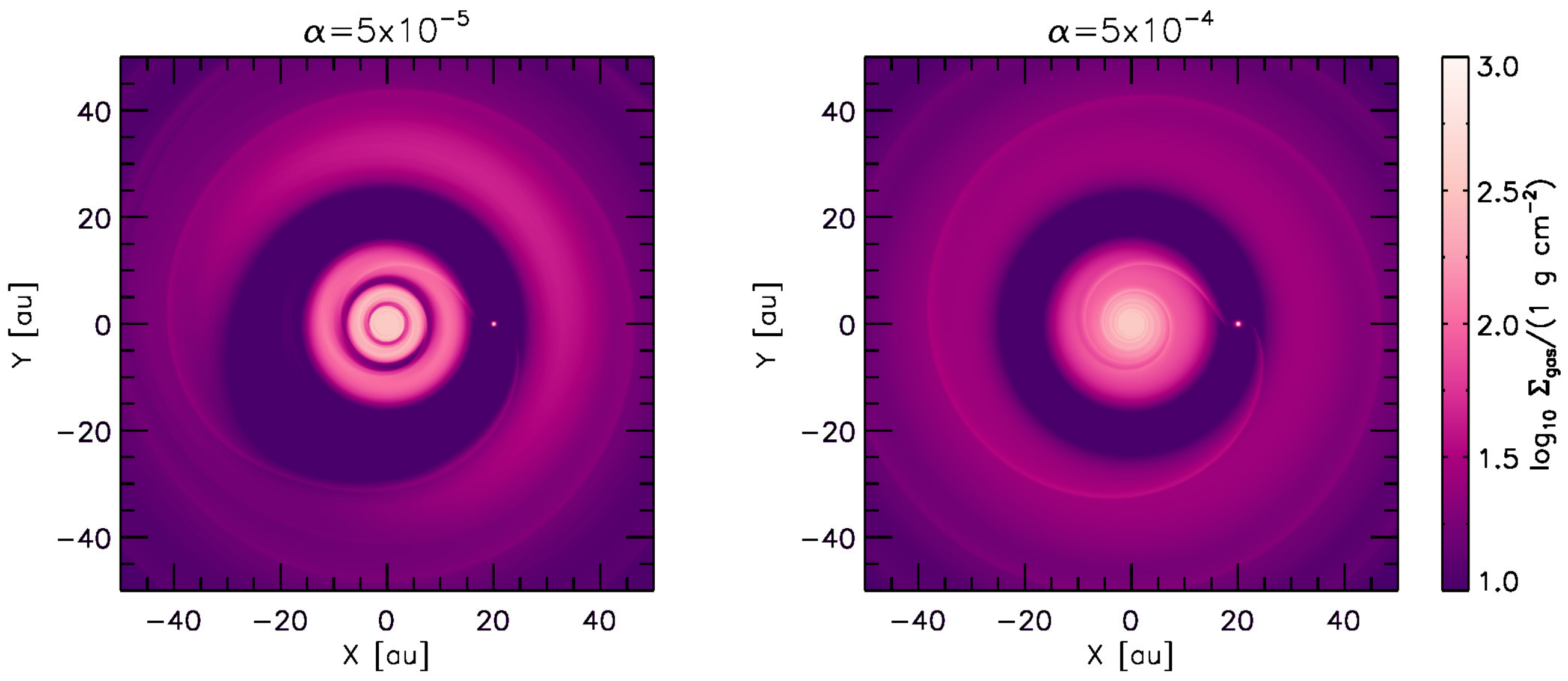}
\caption{The gas surface density distributions from hydrodymanic simulations using (left) $\alpha=5\times10^{-5}$ and (right) $\alpha=5\times10^{-4}$. The Jupiter-mass planet (located at $X=20$~au and $Y=0$~au in the plot) opens three gaps in the low-viscosity disk, while it opens only one gap in the large-viscosity disk. See Figure \ref{fig:opacity} for azimuthally averaged profiles of the gas surface density.
}
\label{fig:hydro}
\end{figure*}

Grains grow via coagulation but their growth can be limited by two barriers.
When the timescale for the fragmentation is shorter than the growth timescale, grains  fragment into smaller bodies rather than further grow: the fragmentation barrier.
Similarly, when the timescale for the radial drift at a certain location is shorter than the growth timescale, the further growth of grains at that location can be limited: the drift barrier.

 While the chain of continuity equations has to be numerically solved to obtain the complete grain size distribution, the maximum size grains can grow under the two barriers can be estimated by comparing the aforementioned timescales \citep[e.g.,][]{birnstiel12}.
In a fragmentation-dominated regime, the maximum size grains can grow ($a_{\rm frag}$) is given by
\begin{equation}
\label{eqn:a_frag}
    a_{\rm frag} \simeq {{\Sigma_{\rm gas}} \over {\pi \rho_s}} {{1-\sqrt{1-4\Lambda_{\rm frag}^2}} \over {\Lambda_{\rm frag}}},
\end{equation}
where 
\begin{equation}
\label{eqn:lambda_frag}
    \Lambda_{\rm frag} \equiv {{1} \over {3 \alpha_{\rm turb}}} \left( {{v_{\rm frag}} \over {c_s}} \right)^2,
\end{equation}
$\rho_s$ is the bulk density of grains, $\alpha_{\rm turb}$ is a parameter characterizing the level of turbulence (defined in a similar manner to the gas viscosity parameter $\alpha$ above), $v_{\rm frag}$ is the fragmentation velocity of grains, and $c_s$ is the sound speed.
In this work, we use $\rho_s = 1.2~{\rm g~cm^{-2}}$ and $v_{\rm frag} = 10~{\rm m~s^{-1}}$.
In a drift-dominated regime, the maximum size grains can grow ($a_{\rm drift}$) is given by
\begin{equation}
\label{eqn:a_drift}
    a_{\rm drift} \simeq {{\Sigma_{\rm gas}} \over {\pi \rho_s}} {{1-\sqrt{1-4\Lambda_{\rm drift}^2}} \over {\Lambda_{\rm drift}}},
\end{equation}
where 
\begin{equation}
\label{eqn:lambda_drift}
    \Lambda_{\rm drift} \equiv {{\Sigma_{\rm grain}} \over {\Sigma_{\rm gas}}} \left( {{v_{\rm Kep}} \over {c_s}} \right)^2 \left| {{d\log P} \over {d\log r}}  \right|^{-1},
\end{equation}
$\Sigma_{\rm grain}$ is the grain surface density, $v_{\rm Kep}$ is the Keplerian velocity, and $P$ is the gas pressure.
Note that the formulae given in Equations (\ref{eqn:a_frag}) and (\ref{eqn:a_drift}) are more general than the ones in literature \citep[c.f.,][]{birnstiel15} to consider small grains having Stokes numbers of $St << 1$ but also large grains having $St \gtrsim 1$ (see also discussions in e.g., \citealt{birnstiel10,pinilla12}).

As can be inferred from Equations (\ref{eqn:a_frag}) and (\ref{eqn:a_drift}), for given disk density and temperature profiles the level of turbulence ($\alpha_{\rm turb}$) determines whether the grain growth is limited by fragmentation or radial drift.
In general, in a disk with strong turbulence the grain growth is limited by fragmentation, whereas in a disk with weak turbulence fragmentation is inefficient and the grain growth is limited by radial drift.
As we will show, however, there exist certain circumstances under which neither fragmentation nor radial drift could limit the growth of grains (see Model 1 below).

\subsection{Model Description}
\label{sec:models}

We consider three models. 
Our fiducial model (hereafter Model 1) uses $\alpha = 5\times10^{-5}$ in the hydrodynamic simulation and $\alpha_{\rm turb} = 5\times10^{-5}$ in the grain evolution calculation.

The second model (hereafter Model 2) uses $\alpha = 5\times10^{-5}$ in the hydrodynamic simulation but an enhanced turbulence parameter $\alpha_{\rm turb} = 10^{-3}$ in the grain evolution calculation.
The physical motivation for separating $\alpha$ and $\alpha_{\rm turb}$ is the following.
In the grain evolution calculation $\alpha_{\rm turb}$ characterizes relative velocities between grains induced by turbulent gas motions which in turn sets the maximum grain size due to fragmentation.
On the other hand, in the hydrodynamic simulation $\alpha$ characterizes the efficiency of viscous transport of gas via any relevant physical processes.
The two parameters therefore do not necessarily characterize the same physical processes.
By purposely separating $\alpha_{\rm turb}$ from $\alpha$ we examine how enhanced fragmentation alone can alter the appearance of the disk with a fixed gas structure, by changing the grain size distribution.

The third model (hereafter Model 3) assumes $\alpha = 5\times10^{-4}$ in the hydrodynamic simulation and $\alpha_{\rm turb} = 5\times10^{-4}$ in the grain evolution calculation.
The purpose of this model is to examine the number of rings and gaps created in a disk with stronger viscous transport.

\section{Simulation Results}
\label{sec:results}

Figure \ref{fig:hydro} presents two-dimensional gas density distributions from hydrodynamic simulations. 
When $\alpha=5\times10^{-5}$ is assumed, the Jupiter-mass planet opens three gaps in the disk. 
The planet opens the primary gap around its orbit at 20~au.
In addition, it opens two gaps in the inner disk at about 8 and 4~au, where the secondary and tertiary spiral arms shock disk gas \citep{bae17}. 
As a result, total three pressure bumps develop in the disk: at $\sim6$~au between the tertiary and secondary gaps, at $\sim 12$~au between the secondary and primary gaps, and at $\sim 33$~au beyond the primary gap. 
When $\alpha=5\times10^{-4}$ is used, on the other hand, the planet opens only the primary gap around its orbit.
This is because the angular momentum transport induced by the secondary and tertiary spiral shocks does not exceed that induced by viscous transport of the disk.

\begin{figure}[t!]
\centering
  \includegraphics[width=0.48\textwidth]{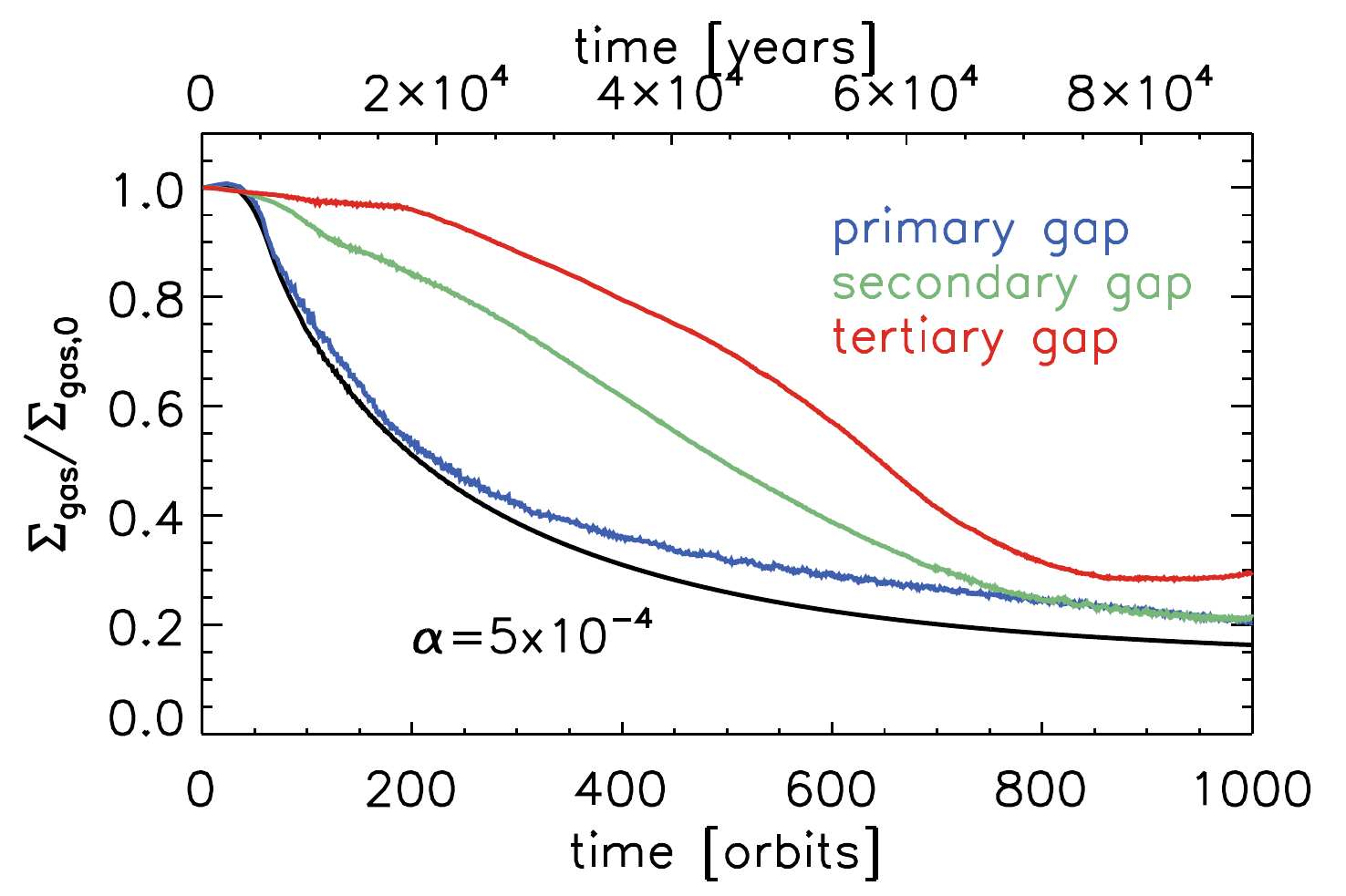}
\caption{The time evolution of gap depths measured in hydrodynamic simulations, defined as the ratio between the gas surface density at the gap center ($\Sigma_{\rm gas}$) to the initial surface density at that location ($\Sigma_{\rm gas,0}$). The color curves show the depth of the three gaps for $\alpha=5\times10^{-5}$ model: (blue) the primary gap at 20~au, (green) the secondary gap at $\sim 8$~au, and (red) the tertiary gap at $\sim 4$~au. The black curve shows the depth of the primary gap (at 20~au) for $\alpha=5\times10^{-4}$ model.}
\label{fig:gap_depth}
\end{figure}

The gap depth in both models reach a quasi-steady state by 1000 planetary orbital times as shown in Figure \ref{fig:gap_depth}.
In the model with $\alpha=5\times10^{-4}$ the primary gap depth is consistent with the analytic estimates and the depths obtained in numerical simulations from literature \citep[e.g.,][]{duffell13,fung14,kanagawa15}.
In case $\alpha=5\times10^{-5}$ is assumed, the primary gap is expected to further deepen over a fraction of the gas viscous timescale \citep[e.g.,][]{fung14}.
However, the drift timescale of grains determining the disk morphology in (sub-)millimeter continuum is much shorter than the gas viscous timescale and, as a result, these grains are drifted toward the adjacent pressure bumps well before the gas viscous timescale is passed as we will show below.

\begin{figure}
\centering
  \includegraphics[height=0.83\textheight]{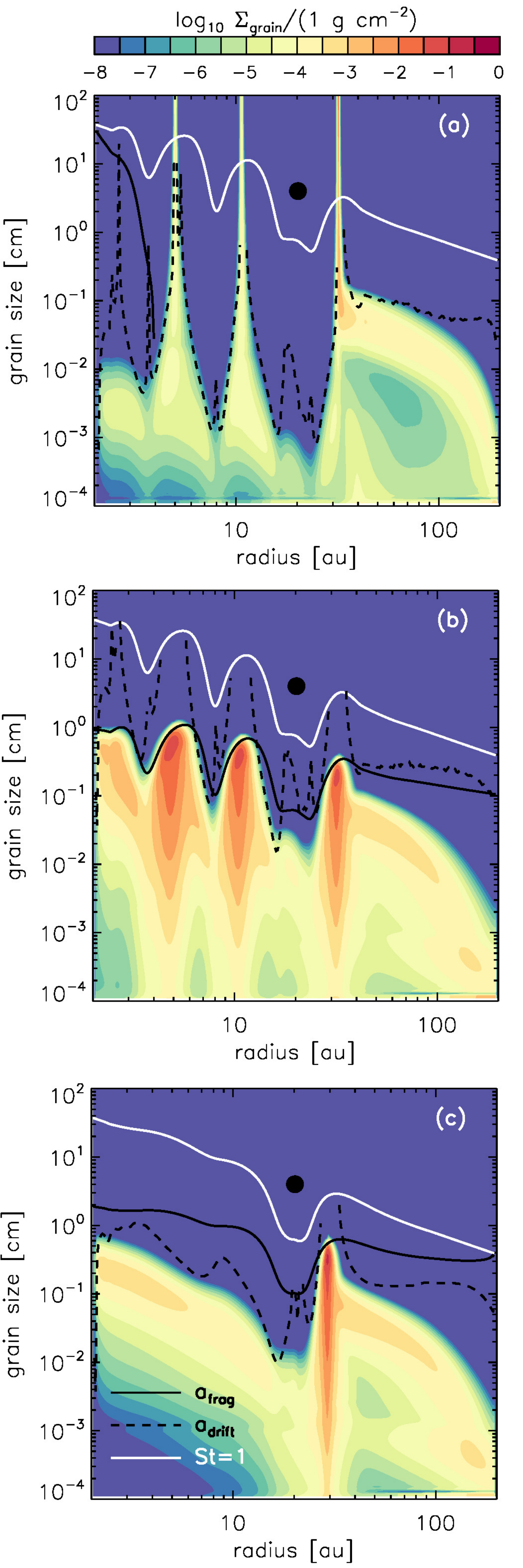}
\caption{Color contours showing the radial distribution of grains with various sizes for (a) Model 1, (b) Model 2, and (c) Model 3. Each horizontal line presents the surface density of the grain ($\Sigma_{\rm grain}$) with the size shown on the y-axis. The white curves show the grain size with Stokes number of unity, while black solid and dashed curves present maximum grain sizes under the fragmentation limit ($a_{\rm frag}$) and the radial drift limit ($a_{\rm drift}$), respectively, calculated using Equations (\ref{eqn:a_frag}) and (\ref{eqn:a_drift}).}
\label{fig:models}
\end{figure}

Figure \ref{fig:models} presents results from the grain evolution calculations after 0.3~Myr.
In Model 1, the maximum grain size is set mainly by the radial drift because the fragmentation is inefficient.
The relative velocity between equal-sized grains due to turbulence can be expressed as $\sqrt{3\alpha_{\rm turb}/ (St + 1/St)} c_s$ \citep[e.g.,][]{cuzzi06,ormel07}.
With a small $\alpha_{\rm turb} = 5\times10^{-5}$ the maximum relative velocity, which  occurs for $St=1$ grains, is smaller than the assumed fragmentation velocity ($v_{\rm frag} = 10~{\rm m~s}^{-1}$) except for inner few au of the disk.
Grains therefore grow with negligible fragmentation until they radially drift toward the pressure bumps.
At the pressure bumps, however, the drift timescale is infinitely long and the radial drift cannot limit the growth of grains.
Within pressure bumps in a disk having sufficiently low turbulence, grains can thus break both fragmentation and radial drift limits and grow beyond meter in size.

Another noticeable feature in this model is that the widths of the dust rings are much narrower than the widths of the gas pressure bumps due to a runaway radial drift (one may compare the width of dust rings in Model 1 with that in Model 2).
Since the radial drift dominates the grain distribution, grains drift toward the adjacent pressure peak as they grow.
The gaps between the pressure peaks lose grains, and this loss leads to an increasingly more efficient radial drift because the growth time scale of grains increases with less grain abundance ($\tau_{\rm grow} \propto \Sigma_{\rm grain}^{-1}$; \citealt{birnstiel12}).
This allows more time for grains to be drifted toward pressure bumps so the drift barrier gradually extends to smaller grain sizes over time.
As a result of this runaway radial drift, grains are collected in a radial region whose width is much narrower than the width of gas pressure bump.

When $\alpha_{\rm turb}$ is increased to $10^{-3}$, the fragmentation limits the growth of grains.
Due to the efficient fragmentation the maximum grain size in the pressure bumps remains at about a few millimeters to a centimeter. 
Also, as grains fragment before they grow to experience rapid radial drift the widths of grain rings remain similar to the widths of pressure bumps.
The widths of grain rings in this efficient fragmentation model are therefore much wider than in Model 1.

In Model 3, the maximum size of grains in the inner disk is determined by the radial drift.
In the pressure bump at the outer gap edge the maximum size of grains is first set by the radial drift, but when grains further drift toward the pressure peak fragmentation determines the maximum size. 

\begin{figure}[t!]
\centering
  \includegraphics[width=0.45\textwidth]{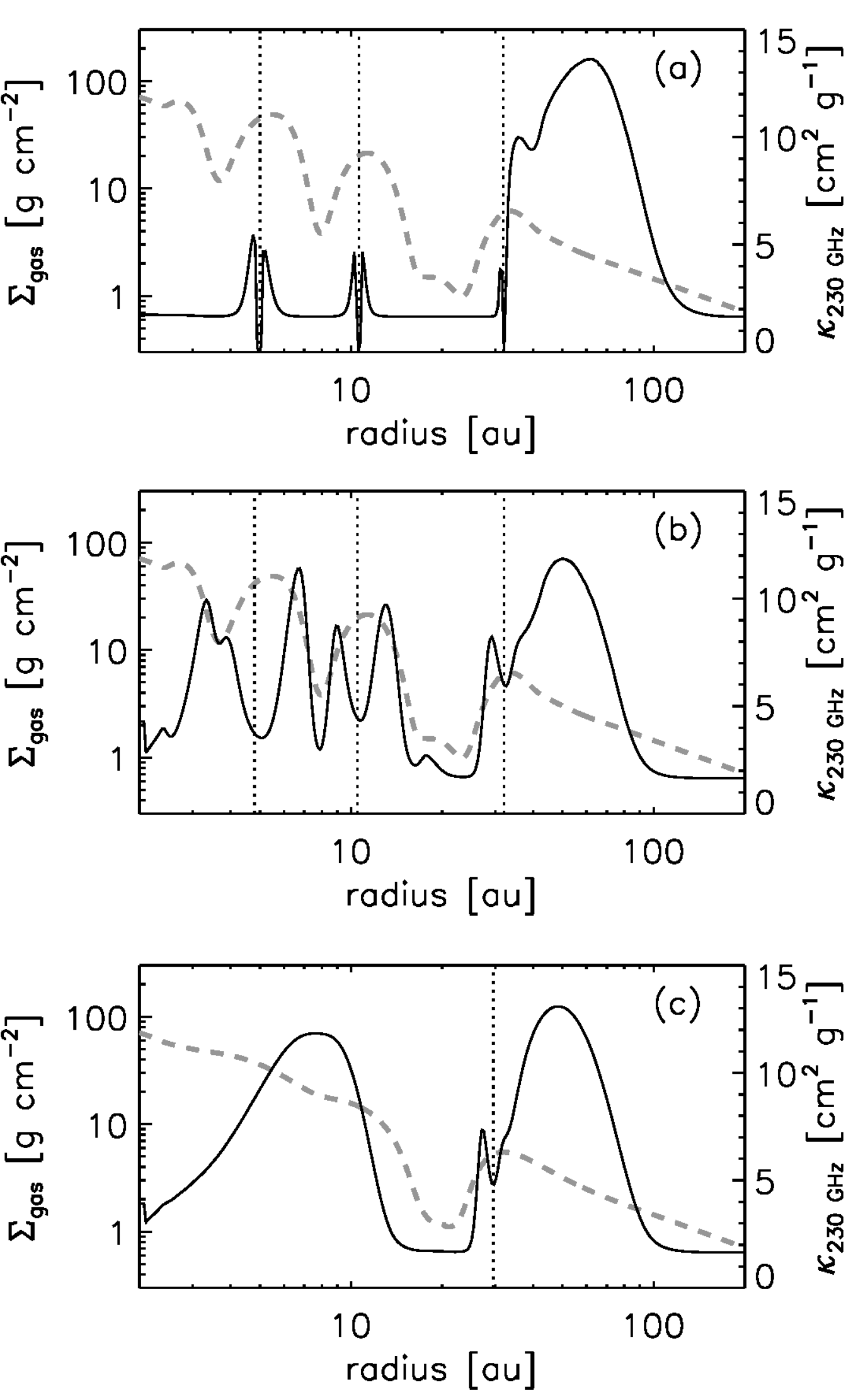}
\caption{The radial distributions of the gas surface density (gray dashed curves, left y-axis) and the continuum opacity at 230~GHz (black solid curves, right y-axis) for (a) Model 1, (b) Model 2, and (c) Model 3. The vertical dotted lines indicate the radial locations where the total grain surface density has a local maximum. Note the radial variations of the continuum opacity across the disks. We adopt the optical constants from \citet{ricci10} to calculate the grain opacity.}
\label{fig:opacity}
\end{figure}

We found in all three models that the grain size distribution in pressure bumps follows a power-law distribution with a slope close to the nominal value of $p=-3.5$, where $dn/da \propto a^{p}$.
However, as can be inferred from Figure \ref{fig:models} the amount of grains dominating the continuum opacity at millimeter wavelengths is very different among the models, but also among different radial locations within each model.
For instance, due to the rapid grain growth millimeter-sized grains are much less abundant in Model 1 than the other models.
Because the total grain mass in Model 1 is dominated by the large grains with sizes of $\gg {\rm mm}$, which do not contribute to opacity at millimeter wavelengths, the opacity within the bumps is very small ($< 0.1~{\rm cm^{2} g^{-1}}$ at 1.3~mm) as shown in Figure \ref{fig:opacity}.

In all three models the continuum opacity changes significantly across the disk.
We thus caution that using a single opacity value when converting the observed continuum flux to the disk's dust mass or dust surface density profile may lead to inaccurate results.

\section{Disk Morphology in Millimeter Continuum}
\label{sec:morphology}

In order to examine the morphology of the model disks in millimeter continuum, we make synthesized images using the grain distribution presented in Figure \ref{fig:models}.
We place disks at 100 pc, with their poles aligned along the line-of-sight (i.e., $0^\circ$ inclination).
We use CASA version 5.1.2\footnote{\url{https://casa.nrao.edu/}} to produce synthesized continuum emission.
Four different antenna configurations C43-9, C43-6, C43-5, and C43-4 are considered, with which at 1.3~mm (230 GHz) we achieve angular resolutions of 0''.025, 0''.13, 0''.24, and 0''.40, respectively.
Thermal noise from the atmosphere and from the antenna receivers is added by setting the {\it thermalnoise} option in the {\it  simobserve} task to {\it tsys-atm}. 
We use on-target integration time of 10 minutes for C43-6, C43-5, and C43-4 antenna configurations and of 1 hour for C43-9 configuration.
The latter is chosen to have comparable spatial resolution and integration time to recent ALMA large program on protoplanetary disks (2016.1.00484.L, PI Andrews, S. M.).
For the four antenna configurations, we obtain rms noise levels around 0.3, 0.1, 0.04, and 0.004~mJy~beam$^{-1}$ for Model 1, and  0.9, 0.3, 0.1, and 0.006~mJy~beam$^{-1}$ for Model 2 and 3.
We assume a uniform dust temperature of 20~K similar to what is typically assumed in continuum surveys of protoplanetary disks \citep[e.g.,][]{andrews05,andrews13}. 
Using the gas temperature for the dust temperature increases the absolute flux of disks and makes the inner disk relatively brighter than the outer disk.
We also examined the disk morphology using dust evolution calculation outputs at various time epochs, but we found no qualitative difference in the morphology.

\begin{figure*}[t]
\centering
  \includegraphics[height=0.9\textheight]{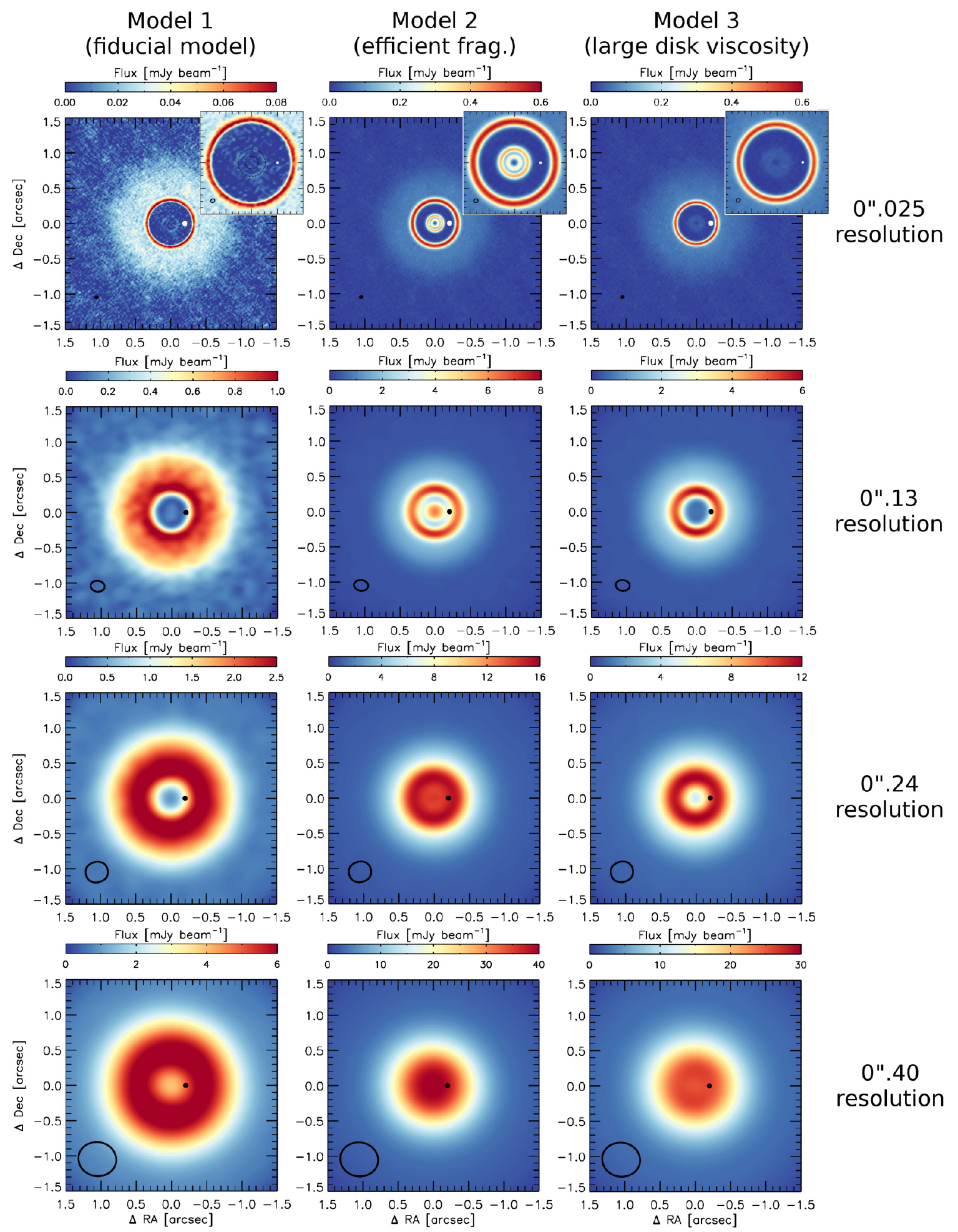}
\caption{Simulated millimeter continuum observations of the three models with (first row) 0''.025, (second row) 0''.13, (third row) 0''.24, and (fourth row) 0''.40 angular resolution. The beam is shown at the lower left part of each panel with a black contour. The planet is located at 0''.2 radial distance from the center (black/white filled circles). The inset panels in the first row present the inner $\pm0''.4$ of the disks. The dotted circles in the inset of highest resolution Model 1 image indicate the locations of the inner two rings.}
\label{fig:synthetic}
\end{figure*}

Figure \ref{fig:synthetic} presents the synthesized images.
The inner two rings have fluxes of 0.003 and 0.006 mJy~beam$^{-1}$ (in the absence of thermal noises), and even with 1 hour of integration the two rings are not detected at the highest resolution in Model 1.
This disk would thus appear to be a transition disk with an inner cavity at all spatial resolution.
The rings have lower fluxes by a factor of $\sim10$ compared with other models due to the low small grain abundance (Figure \ref{fig:models}; see also \citealt{dejuanovelar16}).
As noted earlier, increasing the dust temperature to gas temperature results in higher absolute flux and reveals the inner two rings at $10~\sigma$ level using the highest spatial resolution.
However, even with the enhanced dust temperature the emission from the inner rings are smeared out at lower spatial resolutions, and the disk appears to have an inner cavity.

All the three rings in Model 2 contain sufficient amount of small grains contributing to millimeter continuum.
All the rings are optically thick and the disk appears to have multiple rings and gaps at the highest spatial resolution.
At 0''.13 resolution the inner two rings are not resolved and the disk appears to have a single gap around the planet and a central continuum peak.
At even lower resolutions, the disk appears to be a transition disk with a shallow inner cavity (with a 0''.24 beam) or a full disk without an inner cavity or a gap (with a 0''.40 beam).

One important difference between Model 1 and 2 is that the outer ring in Model 1 is narrower than the outer ring in Model 2 as discussed in the previous section.
When the maximum grain size within pressure bumps is limited by radial drift (Model 1) this can lead a runaway drift toward the pressure peak, resulting in the formation of a very narrow continuum ring (Figure \ref{fig:models}a).
On the other hand, when the maximum grain size within pressure bumps is limited by fragmentation there is no (or little) drift toward the pressure peak and continuum rings have as broad widths as that of gas pressure bumps (Figure \ref{fig:models}b).
A narrow continuum ring width may therefore imply inefficient fragmentation, presumably hinting at low turbulence in the disk.
Alternatively, however, it is possible that a narrow continuum ring width is due to a narrow gas pressure bump width. 
Having high spatial resolution observations that can resolve continuum rings with multiple beams, together with accurate constraints on the radial gas density profile across rings and gaps \citep[e.g.,][]{teague18} and on the disk turbulence \citep[e.g.,][]{flaherty15,teague16}, will allow us to distinguish the two possibilities.

For Model 3 in which only one pressure bump develops, the disk appears to be a transition disk having an inner cavity until the size of the beam becomes comparable to the size of the inner cavity.

\begin{figure}[t!]
\centering
  \includegraphics[width=0.48\textwidth]{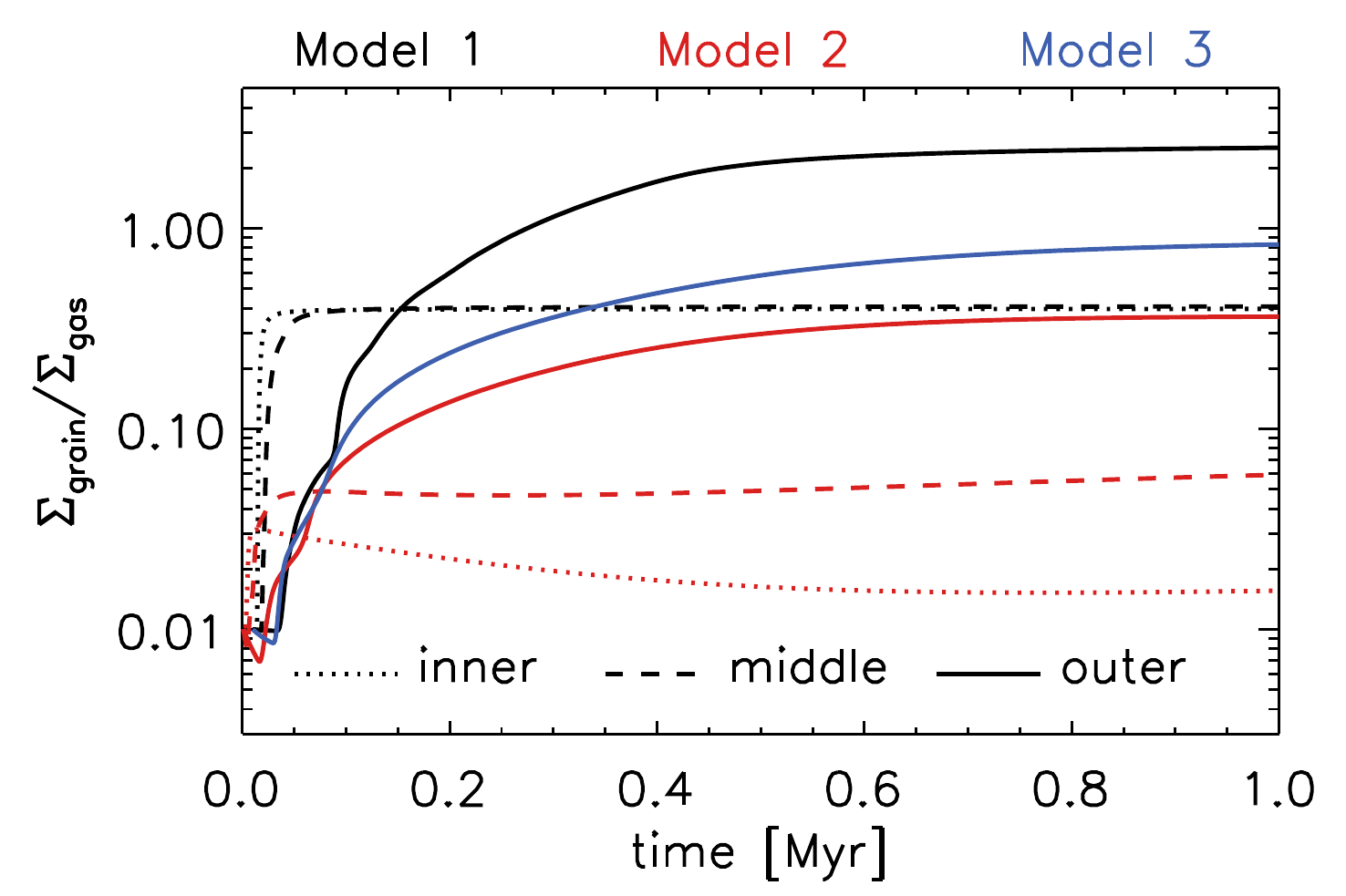}
\caption{The time evolution of dust-to-gas mass ratio $\Sigma_{\rm grain}/\Sigma_{\rm gas}$ at the peak of the dust rings. The black and red curves show the $\Sigma_{\rm grain}/\Sigma_{\rm gas}$ values at the peak of dust rings in Model 1 and 2: (dotted) the inner ring at $\sim5$~au, (dashed) the middle ring at $\sim11$~au, and (solid) the outer ring at $\sim32$~au. The blue solid curve shows the $\Sigma_{\rm grain}/\Sigma_{\rm gas}$ value at the peak of the outer dust ring in Model 3 at $\sim 29$~au.}
\label{fig:gtd}
\end{figure}

As pressure bumps efficiently trap grains they can provide a favorable condition to convert grains to planetesimals and/or protoplanets, whose effects are not included in our dust evolution calculations. 
One such possibility is that the streaming instability is triggered in pressure bumps, converting pebble-sized grains to planetesimals \citep{youdin05,johansen07}.
In Figure \ref{fig:gtd} we present the time evolution of the dust-to-gas mass ratio at the peak of dust rings.
The dust-to-gas mass ratio reaches $\gtrsim 0.1$ within the first $\sim0.2$~Myr of dust evolution, which would be indeed sufficient to trigger the streaming instability \citep{carrera15}.
In addition, when a disk managed to build a Jupiter-mass planet, it is very likely that there exist planetesimals and/or protoplanets throughout the disk.
In such a situation, concentration of grains in pressure bumps can promote the growth of already-existing planetesimals/protoplanets via pebble accretion \citep{johansen10,ormel10}.
In fact, more than 20~Earth-masses of solid particles are collected in the outermost pressure bump in Model 1, potentially facilitating the formation of a second-generation (giant) planet there (\citealt{lyra09,ronnet18}; see also \citealt{pinilla16} for a similar possibility suggested for the outer edge of the dead-zone).

When the (sub-)mm continuum morphology is concerned, a reduction in the amount of grains could decrease the absolute flux of a dust ring.
To test potential effect of the removal of grains via planetesimal-forming/growing processes, we make simulated continuum images after applying a maximum dust-to-gas mass ratio of $\Sigma_{\rm grain}/\Sigma_{\rm gas} = 0.02$.
In practice, at any radius where $\Sigma_{\rm grain}/\Sigma_{\rm gas} > 0.02$ the grain surface density is reduced (uniformly across different dust sizes) to have $\Sigma_{\rm grain}/\Sigma_{\rm gas} = 0.02$.
We chose this particular value because of the following two reasons.
First, $\Sigma_{\rm grain}/\Sigma_{\rm gas} = 0.02$ is about the dust-to-gas surface density ratio below which the streaming instability is unlikely to operate \citep{carrera15}.
Second, it is likely that some fraction of small grains would always remain within the rings even in case streaming instability and/or pebble accretion is in action. 
Numerical simulations show that the conversion rate (in mass) of grains to planetesimals via the streaming instability is less than $50~\%$ \citep{johansen12,simon16}.
Similarly, the efficiency of pebble accretion onto protoplanets, defined as the ratio between the number of pebbles settled to the protoplanet and the total number of integration of pebble's stochastic equation of motion, is $\lesssim 10~\%$ expect at the iceline locations for which the efficiency can be as high as $\sim 50~\%$ \citep{ormel18}.
Furthermore, it is also possible that the collisions among dynamically excited planetesimals in a disk having a giant planet can replenish small grains, as recently proposed for the multi-ringed HD~163296 disk \citep{turrini18}.
So it is reasonable to choose a non-zero $\Sigma_{\rm grain}/\Sigma_{\rm gas}$ value.

\begin{figure*}
\centering
  \includegraphics[width=0.9\textwidth]{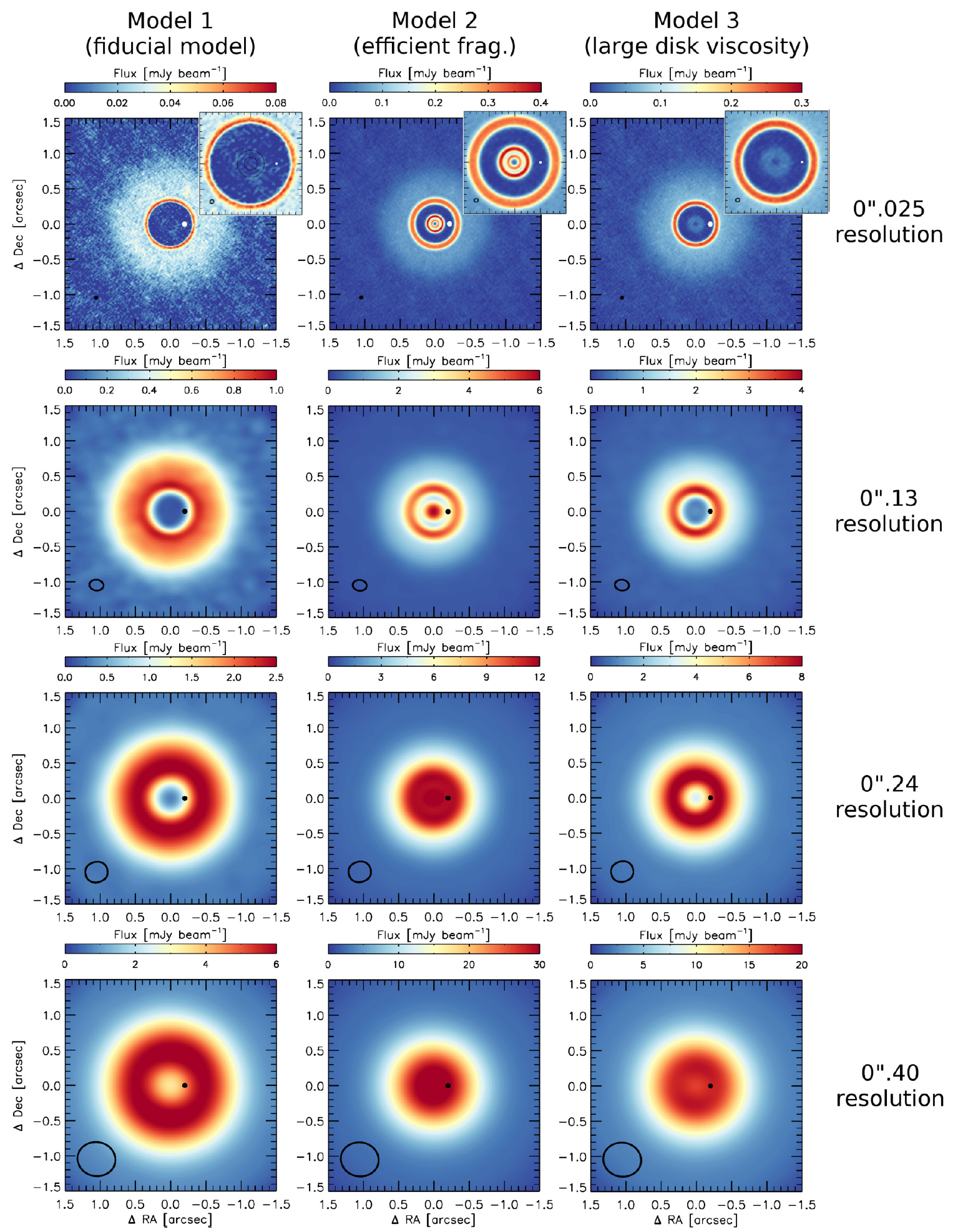}
\caption{Same as Figure \ref{fig:synthetic}, but using the total grain surface at each radius limited to maximum $2~\%$ of the gas surface density to test potential effects of the removal of small grains through planetesimal formation/growth (see text). Note that the color bars in Model 2 and 3 have different ranges than Figure \ref{fig:synthetic}. The absolute flux of the rings has been decreased with the reduced grain amount, but the overall morphology of the disks is unchanged.}
\label{fig:synthetic2}
\end{figure*}

Figure \ref{fig:synthetic2} shows the resulting simulated continuum images. 
The absolute flux of the continuum rings has been decreased with the reduction in grain amount, but it is found that the overall continuum morphology has not been changed because pressure bumps still have sufficient mm-sized grains within the beam.
While this experiment shows that the removal of grains through planetesimal-forming/-growing processes would not significantly alter the overall continuum morphology, further investigations on the efficiency of streaming instability and pebble accretion within pressure bumps and considerations of these processes in dust evolution calculations are warranted to conclude this.

\section{Summary and Outlook}
\label{sec:summary}

Using hydrodynamic simulations, dust evolution models, and synthetic observations, we showed that a Jupiter-mass planet can produce a diverse protoplanetary disk morphology, including a full disk, a transition disk with an inner cavity, a disk with a single gap and a central continuum peak, and a multi-gapped disk.
Such a diversity originates from the level of viscous transport in the disk which determines the number of gaps a planet can open, the grain size distribution set by the radial drift and fragmentation, and the angular resolution used to observe the disk. 

Because disks having the same underlying gas distribution can have different millimeter continuum appearance (Model 1 vs. 2), complementary molecular line observations that can constrain the disk gas distribution are strongly suggested. 
In addition, observations with high spatial resolution and sensitivity are necessary to better understand the true nature of protoplanetary disks.
Finally, searches for localized signatures of planets, including accretion on to planets \citep[e.g.,][]{sallum15}, chemical/kinematic signatures in circumstellar disks at the vicinity of planets \citep{cleeves15,pinte18}, and kinematic/thermal signatures associated with circumplanetary disks \citep{perez15,zhu16,zhu17}, are highly desired to confirm (or disprove) the presence of planets in the disks with substructures and to differentiate possible causes of rings, gaps, and inner cavities.

As illustrated in Figure \ref{fig:substructure}, the hypothesized gap-opening planets locate in an interesting region of the planet mass -- semi-major axis plane, for which we do not have counterpart exoplanets discovered with the current exoplanet detection techniques.
Future observations with 25+ meter class telescopes (e.g., E-ELT, GMT, TMT) will offer unprecedented capabilities to directly detect young, self-luminous planets still embedded in protoplanetary disks, allowing us to have critical insights into the formation and evolution of planets.

\acknowledgments
We thank anonymous referee for quick and helpful reports which have significantly improved the initial manuscript.
J.B. thanks to Richard Teague and Ke Zhang for helpful comments on the initial draft of the paper, to Scott Gaudi for his help with adding the estimated regions of exoplanet detection to Figure \ref{fig:substructure}, to Alycia Weinberger for helpful conversation on future direct imaging observations, and to Andrew Youdin for helpful conversation on the streaming instability. 
J.B. acknowledges support from NASA grant NNX17AE31G and computing resources provided by the NASA High-End Computing (HEC) Program through the NASA Advanced Supercomputing (NAS) Division at Ames Research Center. P.P. acknowledges support by NASA through Hubble Fellowship grant HST-HF2-51380.001-A awarded by the Space Telescope Science Institute, which is operated by the Association of Universities for Research in Astronomy, Inc., for NASA, under contract NAS 5-26555. T.B. acknowledges funding from the European Research Council (ERC) under the European Union's Horizon 2020 research and innovation programme under grant agreement No 714769.

\software{\texttt{CASA} \citep{casa}, \texttt{FARGO3D} \citep{benitez16}}


\begin{thebibliography}{99}

\bibitem[ALMA Partnership et al.(2015)]{alma15} ALMA Partnership, Brogan, C.L., P\'erez, L. M., et al. \ 2015, \apjl, 808, L3 
\bibitem[Andrews et al.(2005)]{andrews05} Andrews, S.~M., \& Williams, J. P.\ 2005, \apj, 631, 1134
\bibitem[Andrews et al.(2013)]{andrews13} Andrews, S.~M., Rosenfeld, K.~A., Kraus, A. L., \& Wilner, D. J.\ 2013, \apj, 771, 129
\bibitem[Andrews et al.(2016)]{andrews16} Andrews, S.~M., Wilner, D.~J., Zhu, Z., et al.\ 2016, \apjl, 820, L40
\bibitem[Avenhaus et al.(2018)]{avenhaus18} Avenhaus, H., Quanz, S.~P., Garufi, A., et al.\ 2018, arXiv:1803.10882
\bibitem[Bae \& Zhu(2018a)]{baezhu18} Bae, J., \& Zhu, Z.\ 2018a, \apj, 859, 118
\bibitem[Bae \& Zhu(2018b)]{baezhu18b} Bae, J., \& Zhu, Z.\ 2018b, \apj, 859, 119
\bibitem[Bae et al.(2016)]{bae16c} Bae, J., Zhu, Z., \& Hartmann, L.\ 2016, \apj, 819, 134 
\bibitem[Bae et al.(2017)]{bae17} Bae, J., Zhu, Z., \& Hartmann, L.\ 2017, \apj, 850, 201
\bibitem[Ben\'itez-Llambay \& Masset(2016)]{benitez16} Ben\'itez-Llambay, P., \& Masset, F. 2016, \apjs, 223, 11
\bibitem[Birnstiel et al.(2015)]{birnstiel15} Birnstiel, T., Andrews, S.~M., Pinilla, P., \& Kama, M.\ 2015, \apjl, 813, L14
\bibitem[Birnstiel et al.(2010)]{birnstiel10} Birnstiel, T., Dullemond, C.~P., \& Brauer, F.\ 2010, \aap, 513, A79
\bibitem[Birnstiel et al.(2012)]{birnstiel12} Birnstiel, T., Klahr, H., \& Ercolano, B.\ 2012, \aap, 539, A148
\bibitem[Carrera et al.(2015)]{carrera15} Carrera, D., Johansen, A., \& Davies, M. B.\ 2015, \aap, 579, A43
\bibitem[Cleeves et al.(2015)]{cleeves15} Cleeves, L. I., Bergin, E. A., \& Harries, T. J.\ 2015, \apj, 807, 2
\bibitem[Cuzzi \& Weidenschilling(2006)]{cuzzi06} Cuzzi, J. N., \& Weidenschilling, S. J. 2006, in Meteorites and the Early Solar System II, ed. D. S. Lauretta \& H. Y. McSween, Jr. (Tucson, AZ: Univ. Arizona Press), 353
\bibitem[de Juan Ovelar et al.(2016)]{dejuanovelar16} de Juan Ovelar, M., Pinilla, P., Min, M., Dominik, C., \& Birnstiel, T. \ 2016, \mnras, 459, L85
\bibitem[Dipierro et al.(2018)]{dipierro18} Dipierro, G., Ricci, L., Per\'ez, L., et al.\ 2018, \mnras, 475, 5296
\bibitem[Dong \& Fung(2017a)]{dongfung17} Dong, R., \& Fung, J.,\ 2017a, \apj, 835, 146
\bibitem[Dong \& Fung(2017b)]{dongfung17b} Dong, R., \& Fung, J.,\ 2017b, \apj, 835, 38
\bibitem[Dong et al.(2015)]{dong15} Dong, R., Zhu, Z., Rafikov, R. R., \& Stone, J. M. \ 2015, \apjl, 809, L5
\bibitem[Dong et al.(2017)]{dong17} Dong, R., Li, S., Chiang, E., \& Li, H. \ 2017, \apj, 843, 127
\bibitem[Duffell \& MacFadyen(2013)]{duffell13} Duffell, P. C., \& MacFadyen, A. I.\ 2013, \apj, 769, 41
\bibitem[Dullemond \& Penzlin(2018)]{dullemond18} Dullemond, C. P., \& Penzlin, A. B. T. \ 2018, \aap, 609, A50
\bibitem[Fedele et al.(2018)]{fedele18} Fedele, D., Tazzari, M., Booth, R., et al.\ 2018, \aap, 610, A24
\bibitem[Flaherty et al.(2015)]{flaherty15} Flaherty, K. M., Hughes, A. M., Rosenfeld, K. A., et al.\ 2015, \apj, 813, 99
\bibitem[Flock et al.(2017)]{flock17} Flock, M., Nelson, R. P., Turner, N. J., et al.\ 2017, \apj, 850, 131
\bibitem[Follette et al.(2017)]{follette17} Follette, K. B., Rameau, J., Dong. R., et al.\ 2017, \aj, 153, 264
\bibitem[Fung et al.(2014)]{fung14} Fung, J., Shi, J.-M., \& Chiang, E. 2014, \apj, 782, 88
\bibitem[Gaudi(2012)]{gaudi12} Gaudi, B. S. \ 2012, ARA\&A, 50, 411
\bibitem[Isella et al.(2016)]{isella16} Isella, A., Guidi, G., Testi, L., et al. 2016, Physical Review Letters, 117, 251101
\bibitem[Jin et al.(2016)]{jin16} Jin, S., Li, S., Isella, A., Li, H., \& Ji, J. 2016, \apj, 818, 76
\bibitem[Johansen et al.(2007)]{johansen07} Johansen, A., Oishi, J. S., Mac Low, M.-M., et al. 2007, Nature, 448, 1022
\bibitem[Johansen \& Lacerda(2010)]{johansen10} Johansen, A., \& Lacerda, P.\ 2010, \mnras, 404, 475
\bibitem[Johansen et al.(2009)]{johansen09} Johansen, A., Youdin, A., \& Klahr, H. 2009, \apj, 697, 1269
\bibitem[Johansen et al.(2012)]{johansen12} Johansen, A., Youdin, A. N., \& Lithwick, Y. 2012, \aap, 537, A125
\bibitem[Kanagawa et al.(2015)]{kanagawa15} Kanagawa, K.~D., Tanaka, H., Muto, T., et al.\ 2015, \mnras, 448, 994
\bibitem[Lin \& Papaloizou(1980)]{lin80} Lin, D. N. C., \& Papaloizou, J.\ 1980, \mnras, 191, 37
\bibitem[Lyra et al.(2009)]{lyra09} Lyra, W., Johansen, A., Klahr, H., \& Piskunov, N.\ 2009, \aap, 493, 1125 
\bibitem[Masset(2000)]{masset00} Masset, F.\ 2000, A\&AS, 141, 165
\bibitem[McMullin et al.(2007)]{casa} McMullin, J.~P., Waters, B., Schiebel, D., et al.\ 2007, Astronomical Data Analysis Software and Systems XVI, 127.
\bibitem[Meru et al.(2017)]{meru17} Meru, F., Juh\'asz, A., Ilee, J. D., et al.\ 2017, \apjl, 839, L24 
\bibitem[Okuzumi et al.(2016)]{okuzumi16} Okuzumi, S., Momose, M., Sirono, S.-i., Kobayashi, H., \& Tanaka, H. \ 2016, \apj, 821, 82 
\bibitem[Ormel \& Cuzzi(2007)]{ormel07} Ormel, C. W., \& Cuzzi, J. N.\ 2017, \aap, 466, 413 
\bibitem[Ormel \& Klahr(2010)]{ormel10} Ormel, C. W., \& Klahr, H. H.\ 2010, \aap, 520, A43 
\bibitem[Ormel \& Liu(2018)]{ormel18} Ormel, C. W., \& Liu, B.\ 2018, arXiv:1803.06150
\bibitem[Perez et al.(2015)]{perez15} Perez, S., Dunhill, A., Casassus, S., et al.\ 2015, \apjl, 811, L5
\bibitem[Pinilla et al.(2012)]{pinilla12} Pinilla, P., Birnstiel, T., Ricci, L., et al. \ 2012, \aap, 538, A114
\bibitem[Pinilla et al.(2016)]{pinilla16} Pinilla, P., Flock, M., de Juan Ovelar, M., \& Birnstiel, T. \ 2016, \aap, 596, A81
\bibitem[Pinte et al.(2018)]{pinte18} Pinte, C., Price, D. J., Menard, S., et al.\ 2018, \apjl, 860, L13
\bibitem[Ricci et al.(2010)]{ricci10} Ricci, L., Testi, L., Natta, A., et al.\ 2010, \aap, 512, A15
\bibitem[Ronnet et al.(2018)]{ronnet18} Ronnet, T., Mousis, O., Crida, A., et al.\ 2018, \aj, 155, 224
\bibitem[Sallum et al.(2015)]{sallum15} Sallum, S., Follette, K. B., Eisner, J. A., et al.\ 2015, Nature, 527, 342
\bibitem[Shakura \& Sunyaev(1973)]{shakura73} Shakura, N.~I., \& Sunyaev, R.~A.\ 1973, \aap, 24, 337 
\bibitem[Sheehan \& Eisner(2018)]{sheehan18} Sheehan, P. D., \& Eisner, J.~A.\ 2018, \apj, 857, 18
\bibitem[Simon et al.(2016)]{simon16} Simon, J. B., Armitage, P. J., Li, R., \& Youdin, A. N.\ 2016, \apj, 822, 55
\bibitem[Takahashi \& Inutsuka(2014)]{takahashi14} Takahashi, S. Z., \& Inutsuka, S.-I. \ 2014, \apj, 794, 55
\bibitem[Teague et al.(2016)]{teague16} Teague, R., Guilloteau, S., Semenov, D., et al. \ 2016, \aap, 592, A49
\bibitem[Teague et al.(2018)]{teague18} Teague, R., Bae, J., Bergin, E. A., Birnstiel, T., \& Foreman-Mackey, D. \ 2018, \apjl, 860, L12
\bibitem[Tsukagoshi et al.(2016)]{tsukagoshi16} Tsukagoshi, T., Nomura, H., Muto, T., et al. \ 2016, \apjl, 829, L35
\bibitem[Turrini et al.(2018)]{turrini18} Turrini, D., Marzari, F., Polychroni, D., \& Testi, L. \ 2018, arXiv:1802.04361
\bibitem[Youdin \& Goodman(2005)]{youdin05} Youdin, A. N., \& Goodman, J.\ 2005, \apj, 620, 459
\bibitem[Zhang et al.(2015)]{zhang15} Zhang, K., Blake, G. A., \& Bergin, E. A. 2015, \apjl, 806, L7
\bibitem[Zhang et al.(2016)]{zhang16} Zhang, K., Bergin, E. A., \& Blake, G. A. 2016, \apjl, 818, L16
\bibitem[Zhu et al.(2017)]{zhu17} Zhu, Z., Andrews, S. M., \& Isella, A. 2017, arXiv:1708.07287
\bibitem[Zhu et al.(2016)]{zhu16} Zhu, Z., Ju, W., \& Stone, J. M. 2016, \apj, 832, 193

\end{thebibliography}
\end{document}